\documentclass[aip,rsi,reprint,amsmath,amssymb,floatfix]{revtex4-1}
\usepackage[british]{babel}
\usepackage{amsmath,amstext,amssymb,graphicx}
\usepackage[utf8]{inputenc}
\usepackage[T1]{fontenc}
\usepackage[unicode,pdfborder={0 0 0},colorlinks=false]{hyperref}
\begin{document}
\title{An Efficient, Movable Single-Particle Detector for Use in Cryogenic Ultra-High Vacuum Environments}
\author{Kaija~Spruck}
\email[]{kaija.spruck@mpi-hd.mpg.de}
\affiliation{Institut für Atom- und Molekülphysik, Justus-Liebig-Universität Gießen, 35392 Gießen, Germany}
\affiliation{Max-Planck-Institut für Kernphysik, 69117 Heidelberg, Germany}
\author{Arno~Becker}
\affiliation{Max-Planck-Institut für Kernphysik, 69117 Heidelberg, Germany}
\author{Florian Fellenberger}
\affiliation{Max-Planck-Institut für Kernphysik, 69117 Heidelberg, Germany}
\author{Manfred~Grieser}
\affiliation{Max-Planck-Institut für Kernphysik, 69117 Heidelberg, Germany}
\author{Robert~von~Hahn}
\affiliation{Max-Planck-Institut für Kernphysik, 69117 Heidelberg, Germany}
\author{Vincent~Klinkhamer}
\affiliation{Max-Planck-Institut für Kernphysik, 69117 Heidelberg, Germany}
\author{Old\v rich Novotný}
\affiliation{Columbia Astrophysics Laboratory, Columbia University, New York, NY 10027, USA}
\author{Stefan~Schippers}
\affiliation{Institut für Atom- und Molekülphysik, Justus-Liebig-Universität Gießen, 35392 Gießen, Germany}
\author{Stephen Vogel}
\affiliation{Max-Planck-Institut für Kernphysik, 69117 Heidelberg, Germany}
\author{Andreas~Wolf}
\affiliation{Max-Planck-Institut für Kernphysik, 69117 Heidelberg, Germany}
\author{Claude~Krantz}
\email[]{claude.krantz@mpi-hd.mpg.de}
\affiliation{Max-Planck-Institut für Kernphysik, 69117 Heidelberg, Germany}
\date{\today}
\begin{abstract}
  A compact, highly efficient single-particle counting detector for ions of keV/u kinetic energy, movable by a long-stroke mechanical translation stage, has been developed at the Max-Planck-Institut für Kernphysik (Max Planck Institute for Nuclear Physics, MPIK). Both, detector and translation mechanics, can operate at ambient temperatures down to $\sim 10$~K and consist fully of ultra-high vacuum (UHV) compatible, high-temperature bakeable and non-magnetic materials. The set-up is designed to meet the technical demands of MPIK's Cryogenic Storage Ring (CSR). We present a series of functional tests that demonstrate full suitability for this application and characterise the set-up with regard to its particle detection efficiency.    
\end{abstract}
\maketitle
\section{Introduction}
\label{introduction}
In the field of ion storage rings dedicated to molecular and atomic physics experiments, the electrostatic Cryogenic Storage Ring (CSR) \cite{RobertNIM,KrantzDR} of the Max Planck Institute for Nuclear Physics (MPIK) in Heidelberg, Germany, bridges the gap between traditional medium-energy magnetic heavy-ion synchrotrons, like TSR \cite{TSRIsolde}, ASTRID \cite{astrid} or CRYRING \cite{cryring}, and smaller low-energy electrostatic storage rings. Most of the latter are off-springs of the ELISA facility \cite{elisa} and target very low ion energies around 20~keV \cite{kek,tmu}. The most advanced designs, like the recently commissioned DESIREE facility in Stockholm \cite{desiree,thomasdesiree}, the future RIKEN storage ring in Saitama \cite{riken}, and also the above mentioned CSR, feature beam guiding vacuum systems that can be cooled to deeply cryogenic temperatures, which results in very low residual gas density and very long storage times\cite{desiree,langectf}.\par   
At a maximum kinetic energy of 300~keV per unit charge, ions stored in the CSR are still low energetic compared to traditional magnetic storage rings. Nevertheless, ions with charge-to-mass ratios up to 1/160~e/u will have storage velocities high enough to allow phase-space cooling using an electron cooler \cite{KrantzDR}. Consequently, CSR will be the first electrostatic storage ring to enable ion-neutral, ion-photon, and ion-electron collision experiments at high precision using cooled, low-emittance stored beams.\par
Collision experiments with stored ions generally produce daughter beams of different ion-optical rigidity compared to their parent, which leave the closed orbit of the storage ring at the next bending element \cite{WolfDR}. The daughter particles can e.g.\ be charge-changed products from electron collisions of highly-charged atomic ions, neutral products from recombination of singly charged cations, or fragments from dissociation of stored molecular or cluster ions. In the case of low-emittance ion beams as stored in an electron cooler ring, movable, single-particle counting detectors of narrow aperture are a well-established way to selectively and sensitively measure the production rates of specific daughters \cite{WolfDR}. The bending elements of the storage ring then act as mass spectrometers that separate different product beams according to their charge-to-mass ratio $q/m$ and, thereby, allow to determine the corresponding reaction cross-sections. In the case of CSR, such a detector has to meet additional technical criteria imposed by the design of the storage ring.\par
A first limitation arises from the characteristic storage energy of CSR. Among its prime experimental targets are singly-charged molecular and weakly-charged atomic cations. Given the maximum kinetic energy of 300~keV per unit charge, limited by the deflection fields of the CSR bending electrodes, the specific kinetic energies of stored ions -- and of daughter beams produced in collision experiments -- can thus be as low as a few keV/u. This can easily result in penetration depths in solids as short as 100~nm \cite{SRIM} and, hence, excludes any detection principle that requires the particles to traverse significant layers of passive material before reaching the detection volume, as is notably the case for surface-barrier semiconductor counters.\par 
Second, the detector has to operate in the rather harsh CSR environment. For the above mentioned advantage of extremely good vacuum conditions, but also in order to enable experiments on rotationally-ground-state molecular ions \cite{ZajfmanDR}, CSR is an all-cryogenic storage ring, where the beam-guiding vacuum chamber as well as all ion beam optics are cooled to $\sim 10$~K by a superfluid helium circuit \cite{RobertNIM}. Both, the detection principle and the mechanics enabling free positioning of the detector at the output of a CSR bending element, must work reliably at this low temperature. In addition, in order to achieve UHV vacuum conditions ($\sim 10^{-11}$~mbar) also in room or intermediate temperature operation, the beam pipe and all beam-facing experimental equipment are required to be bakeable to 250$^\circ$C. In cryogenic mode, a room-temperature equivalent pressure of $~10^{-13}$~mbar is then achieved \cite{langectf}.\par 
Last, in order to enable the CSR to store ions of very low rigidity, e.g. low-energetic protons or light molecular ions like H$_2^+$ or H$_3^+$, disturbances of the ion trajectories induced by stray magnetic fields must be kept at a minimum. All materials used for equipment installed in vicinity of the stored-beam orbit are therefore required to be of very low relative magnetic permeability $\mu_r \le 1.01$.\par
In this paper, we present a movable single-particle detector developed at MPIK for counting of reaction products from ion collision experiments at CSR. In Sect.~\ref{design} we discuss the technical design of the set-up, consisting of the actual particle sensor and the cryogenic translation mechanism. In Sect.~\ref{tests} we characterise the detector with regard to its counting efficiency and reliability of operation in a cryogenic vacuum environment. In Sect.~\ref{commissioning} we report the the first operational test at CSR in room temperature mode. Section~\ref{summary} provides a summary and outlook onto possible future upgrades of the set-up.
\section{Technical Design}
\label{design}
\begin{figure}[tb]
\includegraphics[width=8cm]{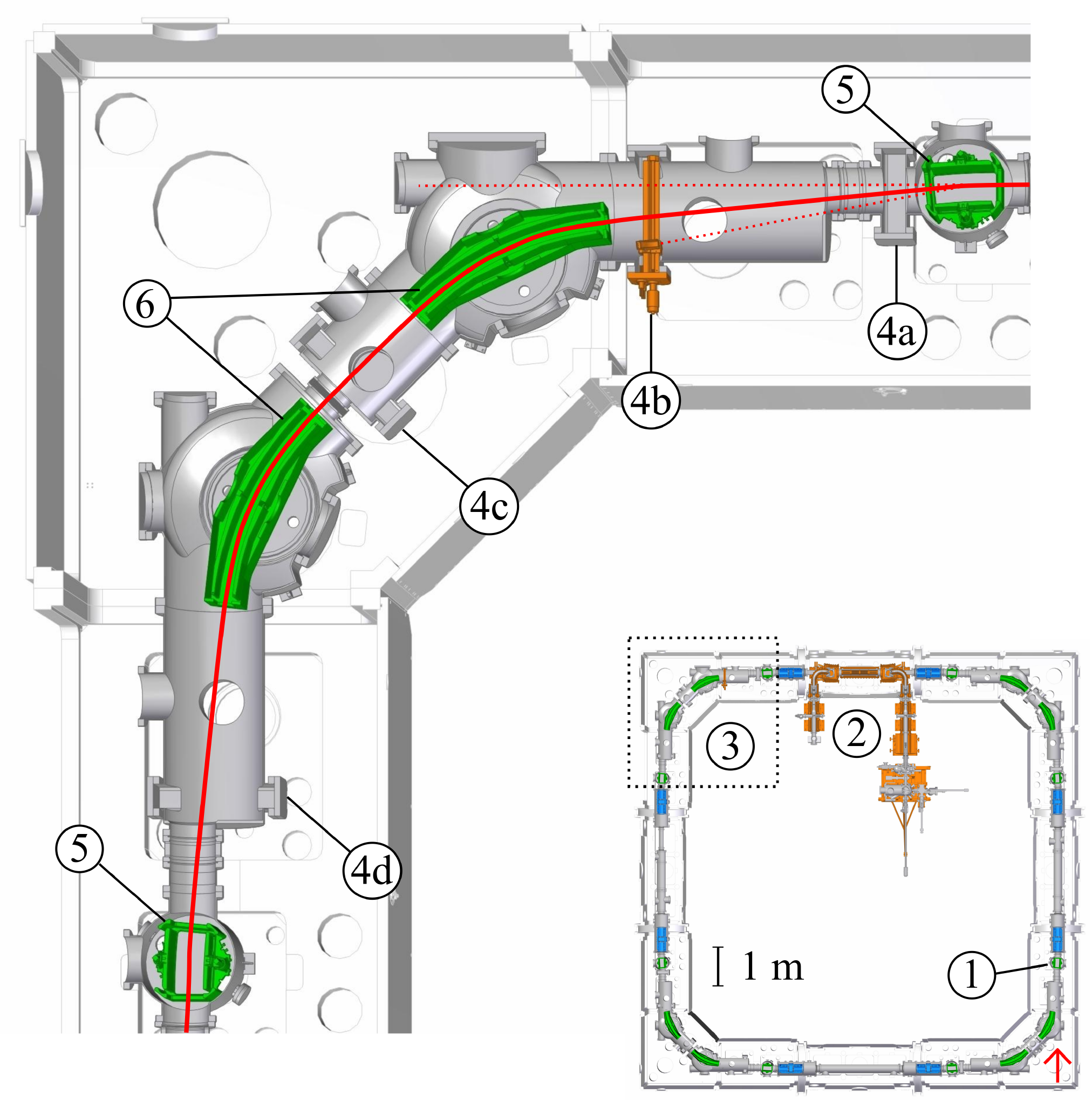}
\caption{\label{csr}(Colour online) Schematic view of the storage ring CSR and close-up of the detector position. Ions are injected by fast switching of a 6$^\circ$ bending electrode (1). Following the electron cooler (2), the third quadrant of the ring (3) features four vacuum ports for the installation of particle detectors (4a--4b). These ports are placed in-between the 6$^\circ$ and 39$^\circ$ bending electrodes (5 and 6), which act as charge-to-mass selectors that separate daughter beams (dotted red lines) from the stored ions (solid red). The here described detector is situated at position (4b).}
\end{figure}

The detector ((4b) in Fig.\ \ref{csr}) is installed behind the 6$^\circ$ bending electrode of the storage ring following the straight section dedicated to the future electron cooler\cite{KrantzDR} of CSR. Limited by the geometry of the beam-guiding vacuum chamber, the detector is, in theory, able to collect daughter particles in the range $ -1.4 \leq \eta \leq +1.1$, where
\begin{equation}
  \label{etaequation}
  \eta = \frac{q_\mathrm{d}/m_\mathrm{d}}{q_\mathrm{p}/m_\mathrm{p}} - 1
\end{equation}
is the relative difference in charge-to-mass ratio $q/m$ of the daughter (index d) compared to the stored parent beam (index p). For the experimentally important case of a positively charged atomic parent beam, the range $-1.0 \leq \eta \leq +1.0$ is most relevant, with the lower and upper boundaries corresponding, respectively, to full neutralisation or ionisation to double charge of the stored parent.\par
The general detector concept was based on the following considerations:
\begin{enumerate}
 \item Detectors relying on surface secondary electron ejection and subsequent multiplication have been shown to achieve very good quantum efficiency for ions in the specific energy range of a few keV/u \cite{rinn}. As noted above, detection principles involving non-sensitive layers that have to be penetrated by the particles before these reach the counting volume are prohibited by the low ion energies at the CSR. 
 \item Micro-channel plates (MCPs) were considered to be the only compact electron multipliers known to work reliably at temperatures of 20~K and below\cite{rosen, kuehnel, roth,schecker}. Especially the so-called `extended dynamic range' (EDR) variants were expected to perform quite well in cryogenic environment thanks to their lower electric resistance. Single-channel electron multipliers (CEMs) were expected to cause problems as their resistance rises exponentially towards low temperatures, causing electron depletion of their single channel wall at high count rates. This assumption turned out to be partly unjustified, as an EDR CEM was successfully tested at 25~K within the test series described in Sect.~\ref{testcryo}. Still, the proven EDR MCP back end was chosen as electron multiplier in the here reported work.
 \item In order to be able to detect a wide spectrum of possible daughter beams in the CSR, the detector must be movable across the diameter of the beam-guiding vacuum chamber, which is approximately 300~mm wide. The standard technique used to move mechanical components across long travel ranges inside UHV chambers are linear actuators, driven from the atmosphere side of the set-up and sealed by long edge-welded bellows. Manufacturers provide no data on their performance at cryogenic temperatures, but excessive compression or expansion of the bellows at low temperatures is explicitly not recommended as embrittlement of the stainless steel welding joints could lead to the formation of vacuum leaks. At the DESIREE project, this problem has been addressed by mounting movable detectors onto sleds that are actuated by stepper motors directly inside the vacuum cryostat. While the magnetic fields of these motors might have been tolerable at CSR as well, provided a certain distance to the stored beam orbit could have been maintained, the electric power used for their operation also leads to significant heating and disturbs the cryogenic environment \cite{thomasdesiree}. Hence, it was decided to move the CSR detector using a cryogenic worm drive operated from the atmosphere side of the set-up. To this end, a rotary UHV feed-through specified for cryogenic temperatures could be obtained commercially, as described later.  
\end{enumerate}
The CSR vacuum system consists of two distinct vacuum vessels: An outer isolation vacuum chamber and -- enclosed by the latter -- an inner cryogenic beam pipe (cf.\ Fig.~\ref{csr}). Rectangular, 80$\times$200 mm$^2$, vacuum ports along the beam line are foreseen for mounting particle detectors off-beam-axis at CSR (cf.\ Figs.\ \ref{csr} and \ref{sensor}). The first operational specimen that is described here is affectionately named `COMPACT', the `COld Movable PArticle CounTer', by its developers. The beam-facing components of the detector, including all mechanical support and all electric connections, were designed to fit entirely through one of the rectangular flanges ((4b) in Fig.~\ref{csr} and (9) in Fig.~\ref{sensor}). Further components in the isolation vacuum chamber are necessary to mechanically and electrically operate the detector from the atmosphere side of the set-up. These had to be designed with low thermal conductance in mind in order not to inflict excessive heat load to the 10-K cold stage of the CSR cryostat.\par
\subsection{Particle Sensor}
\begin{figure}[tb]
\includegraphics[width=8cm]{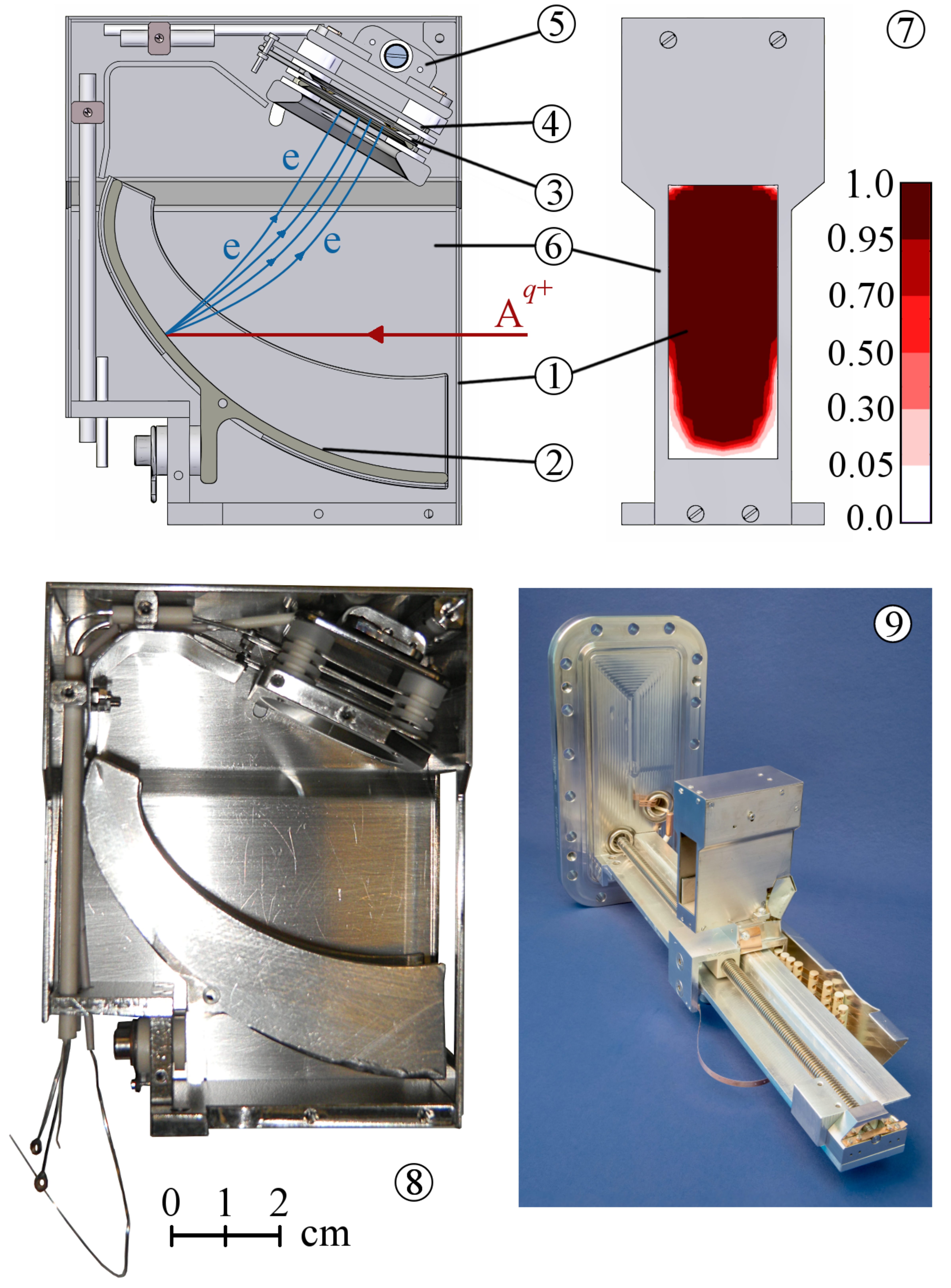}
\caption{\label{sensor}(Colour online) The particle sensor of the detector. The top left and right panels show schematic side and front views, respectively. Particles (denoted by A$^{q+}$) enter the sensor through the 20$\times$50\ mm$^2$ entrance window (1) and impact onto a curved converter cathode (2) (biased at $-400$~V) where they release secondary electrons (e). The latter are collected and multiplied by the MCP stack (3) (biased at $+800$~V and $+2.7$~kV on the input and output sides, respectively). The resulting electron pulse is collected on an anode (4) ($+3.0$~kV) from which the counting signal is decoupled. The MCP-anode stack is equipped with a resistive heating module (5). A grounded housing (6) surrounds the electrodes. The upper right panel (7) shows the simulated secondary electron collection efficiency as a function of the particle impact position in the plane of the entrance window. The lower left (8) and right (9) panels show, respectively, photographic images of the sensor internals and of the fully assembled UHV set-up including the translation stage.}
\end{figure}
The particle sensor is shown in Fig.~\ref{sensor}. It consists of an aluminium secondary-electron converter cathode ((2) in Fig.\ \ref{sensor}), a micro-channel plate and anode stack ((3) and (4) in Fig.\ \ref{sensor}), and a grounded housing encapsulating the latter two ((6) in Fig.\ \ref{sensor}). Incoming heavy particles (labelled A$^{q+}$ in Fig.\ \ref{sensor}) hit the converter cathode, biased at a potential of $-400$~V, and produce secondary electrons. These are accelerated towards the input side of the MCP stack, biased at $+800$~V. The initially slow ($\sim 1$\ eV) secondary electrons thus impinge onto the MCP at a kinetic energy of 1.2~keV. The MCP stack consists of two EDR channel plates (\emph{Photonis}, type 18/12/10/12\,D\,40:1 EDR, MS) of 18~mm useful diameter, assembled in chevron configuration. A potential difference of 1.7 to 2.0~kV can be applied across both plates for electron multiplication. On the output side of the MCPs, a stainless steel anode, biased at $+3.0$~kV with respect to ground, collects the electron bunches.\par
Each primary incident particle generally releases several secondary electrons \cite{baragiola,baroody}. The overall detection probability is hence not limited to the open area ratio of the MCP surface ($\sim$\,60\%), as would be the case for direct heavy particle impact onto the channel plate. In fact, it can reach unity for particles that release many secondaries (cf. Sect.\ \ref{testpulse}). A positive side effect of the secondary-electron detection scheme is that the MCP surface is not exposed to direct impact of high-energy massive particles, which could lead to sputtering damage at high flux densities.\par
Traditionally, charged particle detectors in storage rings have been equipped with apertures a little larger than the expected daughter beam diameter (up to a few mm) and are movable across the storage beam pipe in the two directions transverse to the ion beam axis. Given the difficulties arising from the CSR environment, motion of the sensor along the vertical axis was not considered. Instead, a vertically-elongated sensitive aperture ((1) in Fig.\ \ref{sensor}) of $20\times50$~mm$^2$ acceptance was chosen. A highly efficient ion detector, based on secondary-electron emission, has been described in detail by Rinn et al.\cite{rinn}, and has been used successfully at MPIK for many years. Their design is suitable for a $\sim 1$-cm-diameter circular entrance window. For our purpose, that concept had to be extended towards the above described 20$\times$50~mm$^2$ aperture, which is not only larger in spite of the similar overall size of the device, but also lacks any cylindrical symmetry in the secondary-electron collector optics. Moreover, the sensor has to be enclosed in a grounded housing to prevent disturbances of the ion optics of CSR by stray electric fields. The resulting non-sensitive rim laterally surrounding the detection aperture must be as narrow as possible in order to allow the detector to collect product particles close to the stored-ion orbit without intersecting the latter.\par   
Given these boundary conditions, simulations of the secondary-electron trajectories were performed using the SIMION~8 software package \cite{simion} in order to determine an optimum electrode geometry. A curved conversion cathode, as shown in Fig.~\ref{sensor}, turned out to provide the highest efficiency in transporting the secondary electrons to the MCP stack. As shown in the upper right panel (7) of Fig.~\ref{sensor}, the secondary-electron collection efficiency is $>95$\% over practically the entire area of the sensor aperture. The lateral non-sensitive detector rim could be narrowed down to 2.5~mm ((6) in Fig.~\ref{sensor}).\par
In order to ensure good performance of the electron multiplier stage also at very low ($\sim 10$~K) ambient temperature, a miniature ohmic heating module is included on top of the MCP stack ((5) in Fig.\ \ref{sensor}). Its purpose is to keep the MCPs warmer by 10 to 20~K relative to the experimental environment, thus preventing excessively high channel plate resistance in cold operation of CSR. The heating element is made from an 0.1-mm-thick and 27-cm-long Constantan wire wound around ceramic spacers. It has an electric resistance of $17\ \Omega$ at 300~K, which changes only slightly upon cooling to cryogenic temperature. The various electrodes of the MCP-anode stack are electrically insulated from the heating element by sapphire washers ensuring good thermal coupling. Operational tests of this MCP heating are presented in Sect.\ \ref{testcryo}.   
\subsection{Cryogenic Translation Stage}
\begin{figure}[tb]
\includegraphics[width=8cm]{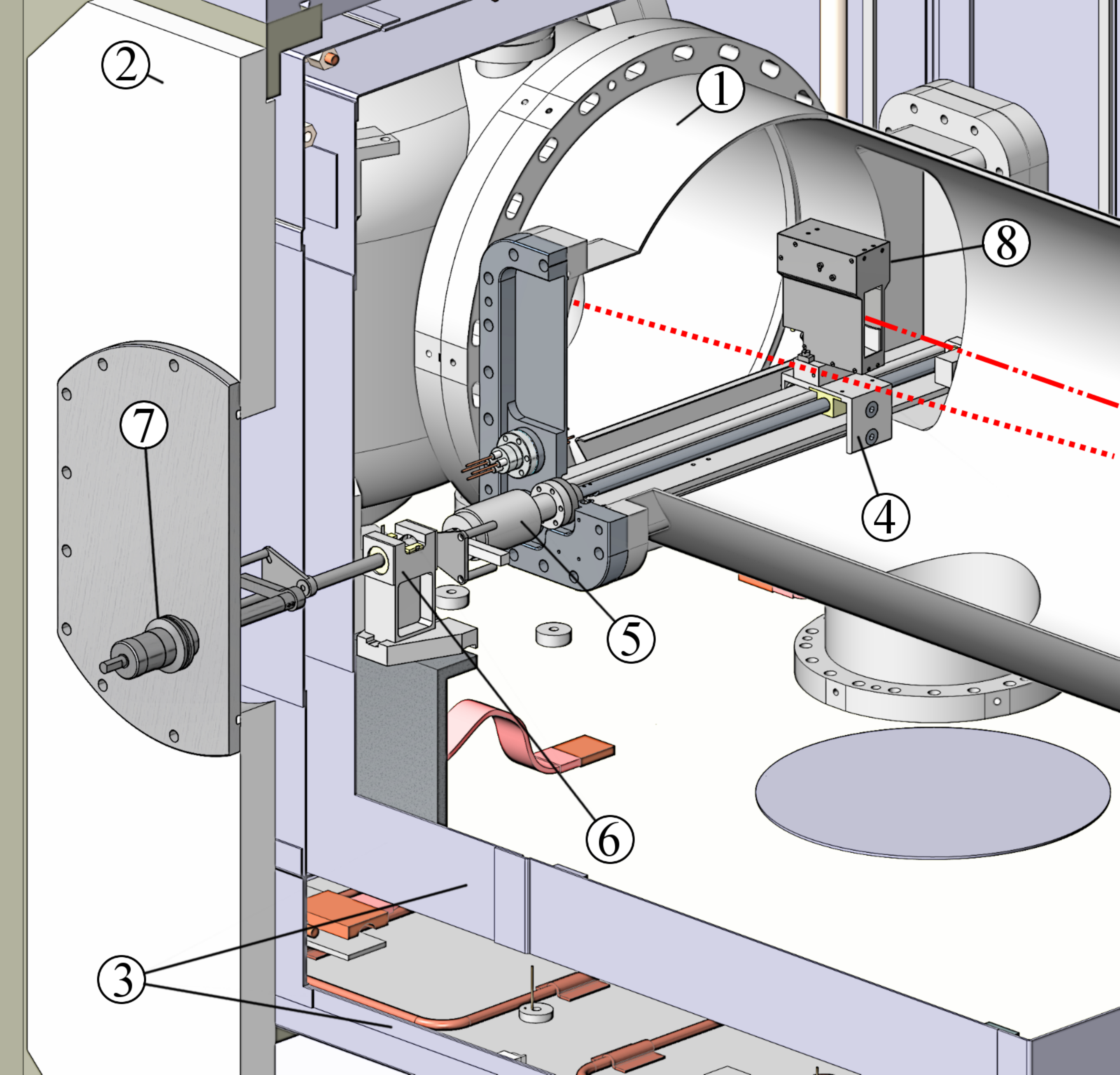}
\caption{\label{CSRdetail}(Colour online) Schematic cut-open view of the detector within CSR. Shown are the 10-K beam pipe (1), the room temperature isolation vacuum chamber (2) and the two thermal radiation shields (3) of the CSR cryostat. The detector mechanics consist of the cryogenic translation stage in the UHV chamber (4), a cryogenic bellow-sealed rotary feed-through (5), a thermally decoupling drive (6) supported by the 40-K stage in the isolation vacuum of the cryostat, and a standard rotary feed-through on the room-temperature chamber wall (7). The movable sensor (8) is connected to the cryogenic high-voltage feed-through via a set of five flexible CuBe2 bands (not shown). The dotted red line represents the closed-orbit trajectory of particles stored in CSR, whereas the dash-dotted red line represents a charged-changed product beam intercepted by the detector.}
\end{figure}  
The translation mechanics of the detector consist of four main components (cf.\ Fig.\ \ref{CSRdetail}): The cryogenic translation stage mounted within the UHV beam pipe of CSR ((4) in Fig.\ \ref{CSRdetail}), a bellow-sealed cryogenic rotary feed-through ((5) in Fig.\ \ref{CSRdetail}) in the experimental chamber wall, a standard high-vacuum rotary feed-through ((7) in Fig.\ \ref{CSRdetail}) on the room-temperature isolation vacuum chamber of the storage ring, and an intermediate rotary drive ((6) in Fig.\ \ref{CSRdetail}) contained within the isolation vacuum vessel of the CSR cryostat.\par
The cryogenic translation stage allows to move the particle sensor across the 300-mm-wide beam-guiding UHV chamber while maintaining secure high-voltage connections between the detector electrodes and electric feed-throughs in the experimental vacuum chamber flange. Motion is induced using a worm drive consisting of a 316L stainless steel Tr10$\times$2-threaded rod, rotating inside a matching nut made of polyether ether ketone (PEEK) and embedded into a carriage supporting the particle sensor. The threaded rod is fixed to the cryogenic rotary feed-through at its one end and loosely contained in an all-ceramic ball bearing at the other. Identical ceramic ball bearings ensure low-friction guidance of the detector carriage along an aluminium rail also attached to the cryogenic vacuum chamber flange. To account for the significantly stronger thermal shrinking of PEEK compared to stainless steel, a very loose fitting ($\sim 0.2$~mm) of the male and female Tr10$\times$2 threads at room temperature had to be chosen. The corresponding loss in positioning precision is irrelevant for the purpose of selecting specific daughter ion beams that are usually separated by several cm. For compatibility with cryogenic operation and UHV requirements, no liquid lubricants are used in any part of the drive. The cryogenic rotary feed-through ((5) in Fig.\ \ref{CSRdetail}) in the experimental vacuum flange is provided off-the-shelf by \emph{Agilent Technologies} (model L6691301, on CF\,16 flange).\par
Electric connections within the beam-guiding vacuum chamber are done via five flexible, 0.1-mm-thick, 4-mm-broad, and approximately 250-mm-long bands of CuBe2 that bend following the motion of the sensor carriage, while being securely fixed to the detector electrodes and heating as well as to the cryogenic high-voltage feed-throughs in the experimental vacuum chamber flange. Spacers made from PEEK ensure electric high-voltage insulation among the CuBe2 bands and towards a grounded Faraday cage surrounding the flexible high voltage lines.\par
\subsection{Isolation Vacuum Mechanics}
Within the isolation vacuum chamber, an intermediate drive ((6) in Fig.\ \ref{CSRdetail}), installed on the 40-K thermal radiation shield ((3) in Fig.\ \ref{CSRdetail}) of the CSR cryostat, provides transmission of rotary motion from the atmosphere side actuator to the cryogenic translation mechanism while ensuring thermal decoupling of the $\sim 10$-K beam-guiding vacuum chamber from the 300-K outer cryostat wall. The 40-K drive contains a spring-loaded sliding bearing with relatively high friction and interconnects to the 10-K and room-temperature rotary feed-throughs via nosings of very loose fitting (approximately $\pm 10^\circ$). By inducing a proper sequence of rotations from the atmosphere side, it is hence always possible to position the 40-K drive in a way that, at rest, it touches neither the room-temperature components nor the 10-K cold stage of the CSR, and hence maintains near-perfect thermal isolation among the different temperature stages of the cryostat. The friction bearing itself is electrically insulating, such that a sensor wire connected to the drive allows the user to check whether it is not touching ground potential at either vacuum chamber.\par
\subsection{Electronics} 
All electronics and power supplies are kept on the atmosphere side of the set-up. Hence electric lines within the isolation vacuum chamber have to connect the room-temperature and cryogenic electric feed-throughs. As electric and thermal conductance strongly correlate for almost all materials, this requires a trade-off between thermal insulation and signal quality. The (static) high-voltage potentials of the converter cathode and of the MCP input and output electrodes are delivered via Kapton-insulated 0.25-mm thick Manganin wires, which have very low thermal conductance, but also high ohmic resistance. For the anode connection, from which the fast counting signal is decoupled, a Kapton-insulated co-axial copper cable (\emph{Allectra} 311-KAP-50-RAD, 50~$\Omega$) is used. In order to limit the resulting thermal load on the 10-K beam pipe of the CSR, the cable is thermally anchored to the intermediate 40-K stage of the cryostat. In addition, the screen of the cable is locally reduced to a single strand at the exit of the 40-K volume. In this way the heat conductivity of the screen is greatly reduced while galvanic contact is maintained. Still, due to the high thermal conductance of copper, the resulting detector-related heat input to the cryogenic beam line is dominated by the co-axial anode connection. It can be estimated to $25(5)$\,mW using tabulated heat conductivity values \cite{cryoengeneering}.\par
The high-voltage potentials for the converter, anode, and MCP input electrodes are provided by standard low-ripple power supplies (\emph{ISEG} NHQ 204M). The gain voltage is generated by a floating high-voltage supply (\emph{Spellman} MCP Series) stacked onto the MCP input potential. The electron current pulses arriving at the anode are separated from the high potential by a decoupling capacitor directly outside the isolation vacuum feed-through and driven into a 50-$\Omega$ current-sensitive fast front-end amplifier (\emph{ORTEC} VT120A) with gain factor 200. Normally no further amplification stage is needed before discrimination of the pre-amplified pulses.      
\section{Functional Tests\label{tests}}
In order to ensure suitability for operation at CSR, a series of tests of the detector set-up were performed. The assembled sensor and translation stage were baked to 250$^\circ$C without resulting damage. Functional tests of the sensor with regard to signal quality and counting efficiency were performed. Furthermore, electrical and mechanical performance of the set-up at low temperature was tested.   
\subsection{Counting Efficiency\label{testpulse}}
The detection principle based on conversion of the primary incident particle into several secondary electrons, together with good collection efficiency of the secondaries by the positively biased MCP stack, suggests good overall particle counting performance of the set-up. This section provides a theoretical and experimental description of the detection efficiency of the sensor.\par
\subsubsection{Theory}
We assume that the entire sensitive aperture of the sensor is irradiated by a homogeneous flux of primary particles. Let $\gamma$ be the average secondary electron yield per primary heavy particle in a certain spot of the converter cathode, and $\epsilon_\mathrm{c}$ the likelihood for any given secondary electron to reach the MCP input surface. Both $\gamma$ and $\epsilon_\mathrm{c}$ may vary spatially across the area of the converter cathode. It is well established that the number $n$ of secondary electrons arriving at the MCP can then be described by a Pólya distribution\cite{prescott,dietz,lakits} $W_n(\tilde{\gamma},b)$ given by  
\begin{equation}
 W_n(\tilde{\gamma},b) = \frac{\tilde{\gamma}^n}{n!}(1+b\,\tilde{\gamma})^{-n-1/b}\ \prod_{j=0}^{n-1} (1 + j\,b)\ , 
 \label{polya}
\end{equation}
where $\tilde{\gamma}$ is the mean of the secondary electron yield $\gamma$ of the cathode, weighted with the probability $\epsilon_\mathrm{c}$ for a given electron to reach the MCP input surface, i.e.\ $\tilde{\gamma} = \left\langle \gamma\epsilon_\mathrm{c} \right\rangle$. The dimensionless parameter $b$ with $0\leq b \leq 1$ is the relative variance\cite{prescott} of the product $\gamma \epsilon_\mathrm{c}$.\par
We assume further that each electron impinging on the MCP has a probability $\epsilon_\mathrm{M}$ of triggering an avalanche in a channel. The total number $k$ of MCP cascades resulting from one incident primary particle is then distributed according to
\begin{eqnarray}
 P_k  & = & \sum_{n=k}^\infty B(k\,\vert\,\epsilon_\mathrm{M}, n)\,W_n(\tilde{\gamma},b) \\
      & = & \sum_{n=k}^\infty \binom{n}{k}\epsilon_\mathrm{M}^k(1-\epsilon_\mathrm{M})^{n-k} \,W_n(\tilde{\gamma},b)\ ,
      \label{pulseheightpolya}
\end{eqnarray}   
since the binomial distribution $B$ gives the probability for $k$ electrons out of $n$ triggering an MCP cascade at individual triggering probability $\epsilon_\mathrm{M}$, and as the number $n$ of impinging secondaries is itself Pólya-distributed according to Eq.~(\ref{polya}).\par 
A large variance of the product $\gamma\epsilon_\mathrm{c}$ is reflected by a high value of $b$. In the limit $b\rightarrow 1$, $W_n$ converges to the geometric distribution. For zero variance, i.e.\ for $b = 0$, $W_n$ becomes a Poisson distribution \cite{prescott} of mean $\tilde{\gamma}$. Hence, if only a sufficiently small, specific spot of the cathode is irradiated by primary particles, Eq.\ (\ref{pulseheightpolya}) simplifies to  
\begin{eqnarray}
 P_k^\prime  = \lim\limits_{b\rightarrow 0} P_k & =  & \sum_{n=k}^\infty \binom{n}{k}\epsilon_\mathrm{M}^k(1-\epsilon_\mathrm{M})^{n-k}\frac{\tilde{\gamma}^n}{n!}e^{-\tilde{\gamma}} \\
      & = & \frac{\left( \epsilon_\mathrm{M}\epsilon_\mathrm{c}\gamma \right)^k}{k!}e^{-\epsilon_\mathrm{M}\epsilon_\mathrm{c}\gamma} \ ,
      \label{pulseheight}
\end{eqnarray}
where we have set $\tilde{\gamma} = \left\langle \gamma\epsilon_\mathrm{c}\right\rangle = \gamma\epsilon_\mathrm{c}$ as, in contrast to Eq.~(\ref{pulseheightpolya}), $\gamma$ and $\epsilon_\mathrm{c}$ can now be considered constant over the cathode area of interest.\par
It is worth noting that the $k$ secondary electrons that multiply successfully should be assumed to impinge into \emph{different} MCP channels. In contrast to a single-channel electron multiplier, whose gain is damped by the space-charge limit of the electron avalanche for high numbers of impinging electrons, the output signal amplitude of the MCP is therefore proportional to $k$.\par
\begin{figure}[tb]
\includegraphics[width=8cm]{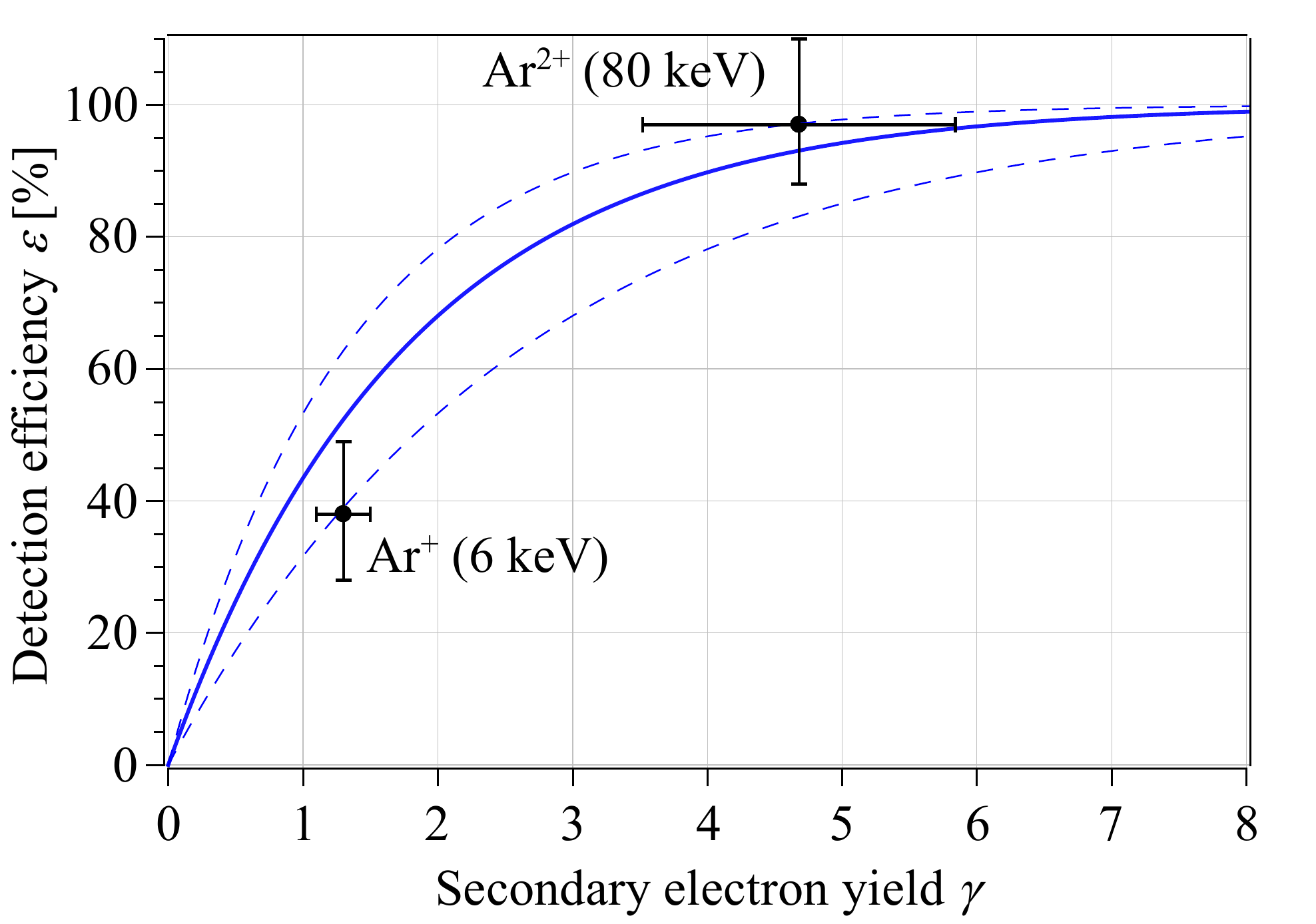}
\caption{\label{effiplot}(Colour online) Theoretical detection efficiency $\varepsilon$ (solid curve) expected according to Eq.~(\ref{efficiency}) as a function of the converter secondary electron yield $\gamma$ for fixed values $\epsilon_\mathrm{M}=0.60$ and $\epsilon_\mathrm{c}=0.95$. The dashed lines represent the systematic error based on an estimated relative uncertainty of 33\% on $\epsilon_\mathrm{M}$ (see text). The filled circles represent the measured values $\varepsilon(\gamma)$ with their experimental uncertainties.}
\end{figure}

For a detector irradiated by a product beam from an electron-cooled, low-emittance ion beam \cite{dimacool}, the irradiated area of the converter cathode can be smaller than $0.1$~mm$^2$. Hence, under the assumption that the spatial variation of $\gamma\epsilon_\mathrm{c}$ is negligible across this characteristic beam size, Eq.~(\ref{pulseheight}) is applicable. Under the further assumption that a primary particle incident onto the sensor is successfully detected as long as it triggers \emph{at least one} electron avalanche, the overall detection efficiency $\varepsilon$ is then given by 
\begin{equation}
  \label{efficiency}
  \varepsilon = 1 - P_0^\prime = 1 - e^{-\epsilon_\mathrm{M}\epsilon_\mathrm{c}\gamma}\ . 
\end{equation}
As $\epsilon_\mathrm{c} \ge 0.95$ over practically the entire detector aperture according to simulations (cf.\ Sect.~\ref{design} and Fig.\ \ref{sensor}), $\varepsilon$ depends mainly on the local secondary electron yield $\gamma$ and the MCP efficiency $\epsilon_\mathrm{M}$ for 1.2-keV electrons. Both numbers are difficult to predict theoretically. It is common practice to use the open area ratio of the MCP of 0.6 as estimate for $\epsilon_\mathrm{M}$. The few direct measurements for electron energies around 1~keV that have been published largely confirm this value, but come with quite large uncertainties \cite{fraser,wiza,galanti}. In the following, we thus assume $\epsilon_\mathrm{M} = 0.6\,(2)$. The value of $\gamma$ is even harder to estimate. While electron ejection from pure metal targets can be modelled quite reliably, it is known that, in practice, the yield of secondaries depends not only on the velocity of the heavy projectile and the material of the cathode (Al), but also on the chemical composition and thickness of any oxide or other adsorption layers that may have formed on the Al surface \cite{baragiola, baroody, dietz,moshammer}. These latter parameters are difficult to control. It is however safe to assume that, in average, several electrons are ejected if the velocity of the incident particle is high enough. From Eq.~(\ref{efficiency}) one calculates that starting from an electron yield of $\gamma=5$ the overall detection efficiency $\varepsilon$ should be greater than 95\%, as shown in Fig.~\ref{effiplot}.
\subsubsection{Experiment}
The above theoretical model for the detection efficiency was tested empirically using the `pinhole' Faraday cup technique that has been described in detail by other authors \cite{fricke,rinn}.\par
A beam of $^{40}$Ar$^{2+}$ ions was produced from a Penning-type ion source, momentum-selected in a bending dipole magnet, and accelerated to a total kinetic energy of 80~keV (2~keV/u) using the CSR injector platform \cite{KrantzDR}. After focusing and collimation to a cross-section of 2.6$\times$2.7~mm$^2$, the remaining beam current of approximately 0.5~nA was dumped onto a Faraday cup. A circular pinhole aperture of 10.5\,(5)~$\mu$m diameter at the bottom of the cup allowed a very small ($\sim 0.001$\%) portion of the ion beam to pass and impinge onto the sensor. In this way, a rate of Ar$^{2+}$ ions suitable for single particle counting (a few $10^4$~s$^{-1}$) arrived at the detector, while the effect of the pinhole aperture on the overall ion current measured at the Faraday cup was negligible. The pinhole also made sure that only a very small spot on the conversion cathode was irradiated by ions, and its position was approximately centred onto the sensitive area of the detector.\par
The ion beam was then steered across the pinhole aperture in two dimensions ($x$ and $y$) using small dipole magnets following the collimator. The Faraday cup aperture was much larger than the beam cross section. Hence the cup always measured the full beam current $I$, independent on the steerer settings. The rate of particles $r_{ij}$ detected by the sensor after the pinhole, and normalised to the corresponding beam current $I_{ij}$, was measured for a matrix of beam deflections $(x_i,y_j)$. The experimental detection  efficiency can then be calculated\cite{rinn} as
\begin{equation}
  \label{rinnformula}
  \varepsilon = \frac{q\,e\,A}{a}\,\sum_{i,j} \frac{r_{ij}}{I_{ij}}\ ,
\end{equation}
where $a$ is the area of the pinhole aperture, $I_{ij}$ the measured ion current in the cup, $e$ the elementary charge, $q$ the charge state of the ions, and $A$ the mesh area of the scanning matrix  $(x_i,y_j)$:
\begin{equation}
  A = (x_{i+1} - x_i)\times (y_{j+1} - y_j)\ .
\end{equation}
\par
In this way, an experimental detection efficiency for 80-keV Ar$^{2+}$ of $\varepsilon_{\mathrm{Ar}^{2+}(80\,\mathrm{keV})} = 97\,(3)\,\%$ was obtained, and could be reproduced several times within the given uncertainty. The 3\% statistical error bar stems mainly from the ion current measurement $I_{ij}$ at the Faraday cup. Larger systematic uncertainties arise from determination of $a$: Direct microscopic measurement of the pinhole diameter ($\pm 0.5\,\mu$m) leads to a relative uncertainty of $\pm 9\,\%$. Due to the thickness of the pinhole aperture of 12.7~$\mu$m, the angle of incidence of the ion beam onto the cup, estimated to $0 \pm 3^\circ$, leads to an additional, asymmetric error bar of $^{+0}_{-8}\,\%$ on $a$, and, hence, of $^{+8}_{-0}\,\%$ on the measured detection efficiency $\varepsilon$.\par
A second, similar, experiment was performed using a very slow beam of Ar$^+$ of 6~keV energy (150~eV/u). As the CSR injector was then not available, another, simpler, ion source and extraction beam line was used. In the same way as described above, a detection efficiency $\varepsilon_{\mathrm{Ar}^{+}(6\,\mathrm{keV})} = 38\,(9)\,\%$ was obtained. The lesser statistical precision of this experiment results mainly from the fact that the beam delivered by the small ion source suffered from lower intensity and worse collimation. The former led to increased uncertainties of $I_{ij}$, the latter to a larger error bar on $r_{ij}$, as background count rate from stray ions in the vacuum chamber had to be subtracted.\par
The main benefit of the experiment with 6-keV ions was that a direct measurement of the secondary electron yield $\gamma$ could be performed: The aperture at the bottom of the Faraday cup was enhanced such as to allow the full ion beam of 0.1 to 1~nA to impinge on the converter cathode. The resulting total (impinging ions plus emitted electrons) current flowing at the cathode could be directly measured by a floating nano-amperemeter. By comparison to the independently measured primary ion current a value of the secondary electron yield of $\gamma_{\mathrm{Ar}^+(6\,\mathrm{keV})} = 1.3\,(2)$ could be determined. As noted above, the anode pulse heights are proportional to the number $k$ of secondary electrons that are successfully multiplied in the MCP, and are hence distributed according to $P^\prime_k$ from Eq.~(\ref{pulseheight}), whose mean is $\gamma\,\epsilon_\mathrm{M}\epsilon_\mathrm{c}$. The ratio of mean anode pulse heights observed in the experiments on Ar$^{2+}$ (80~keV) and Ar$^{+}$ (6~keV) suggested that the secondary electron yield for the former was higher by a factor 3.6\,(7), i.e.\ $\gamma_{\mathrm{Ar^{2+}(80\,\mathrm{keV})}} = 4.7\,(1.2)$. The observed secondary electron yields are lower than what has been reported for Argon impacting onto adsorption-contaminated Al surfaces \cite{schackert,collins}, but also significantly higher than the values published for atomically clean Al targets \cite{baragiola,magnuson}.\par
The results of both measurements of $\varepsilon(\gamma)$ are shown in Fig.\ \ref{effiplot} with error bars representing the quadrature sums of their statistical and systematic uncertainties. While the latter are quite large, the agreement with the simple model from Eqs.\ (\ref{pulseheight}) and (\ref{efficiency}) is reasonable. We note that for the higher energetic 80~keV (2~keV/u) Ar ions, a very good overall detection efficiency of practically 1 is observed. In CSR operation, with typical ion energies of a few keV/u, a similar performance can be expected.   
\subsection{Cryogenic Operation\label{testcryo}}
Compatibility with cryogenic operation was tested, separately for the particle sensor and translation stage, using a test cryostat equipped with a \emph{Leybold} RPK-800 dual-stage pressurised helium refrigerator. A thermal shield attached to the first cold head stage cools down to $\sim 40$~K and prevents radiative heat input onto the experimental platform. The latter consists of a 400$\times$80~mm$^2$ copper support fixed to the second stage of the cold head. It is sufficiently spacious to hold the detector translation stage or the particle sensor, and cools to $\sim 17$~K without heat load. The set-up is evacuated to a base pressure of $1\times 10^{-6}$~mbar by a turbo-molecular pump prior to cooling down.
\subsubsection{MCP performance}
The particle sensor was cooled to a temperature of approximately 20~K while being irradiated by an $^{241}$Am $\alpha$ emitter. The latter had an activity of 3.4~MBq, but was mounted in the room-temperature part of the vacuum chamber and collimated in such a way that $\alpha$ particles entered the detection aperture at an estimated rate of $\sim 4\times 10^3\ \mathrm{s}^{-1}$. Upon impact onto the converter cathode, the 5.4-MeV $\alpha$ particles generated secondary electrons which were multiplied in the MCPs. After decoupling and amplification, as described in Sect.~\ref{design}, the resulting anode current pulses were converted into digital counts using a linear discriminator. The discrimination threshold was deliberately set high to make sure that no electric noise, that could have occurred during the experiment, would be misinterpreted as particle hits.\par
\begin{figure}[tb]
\includegraphics[width=8cm]{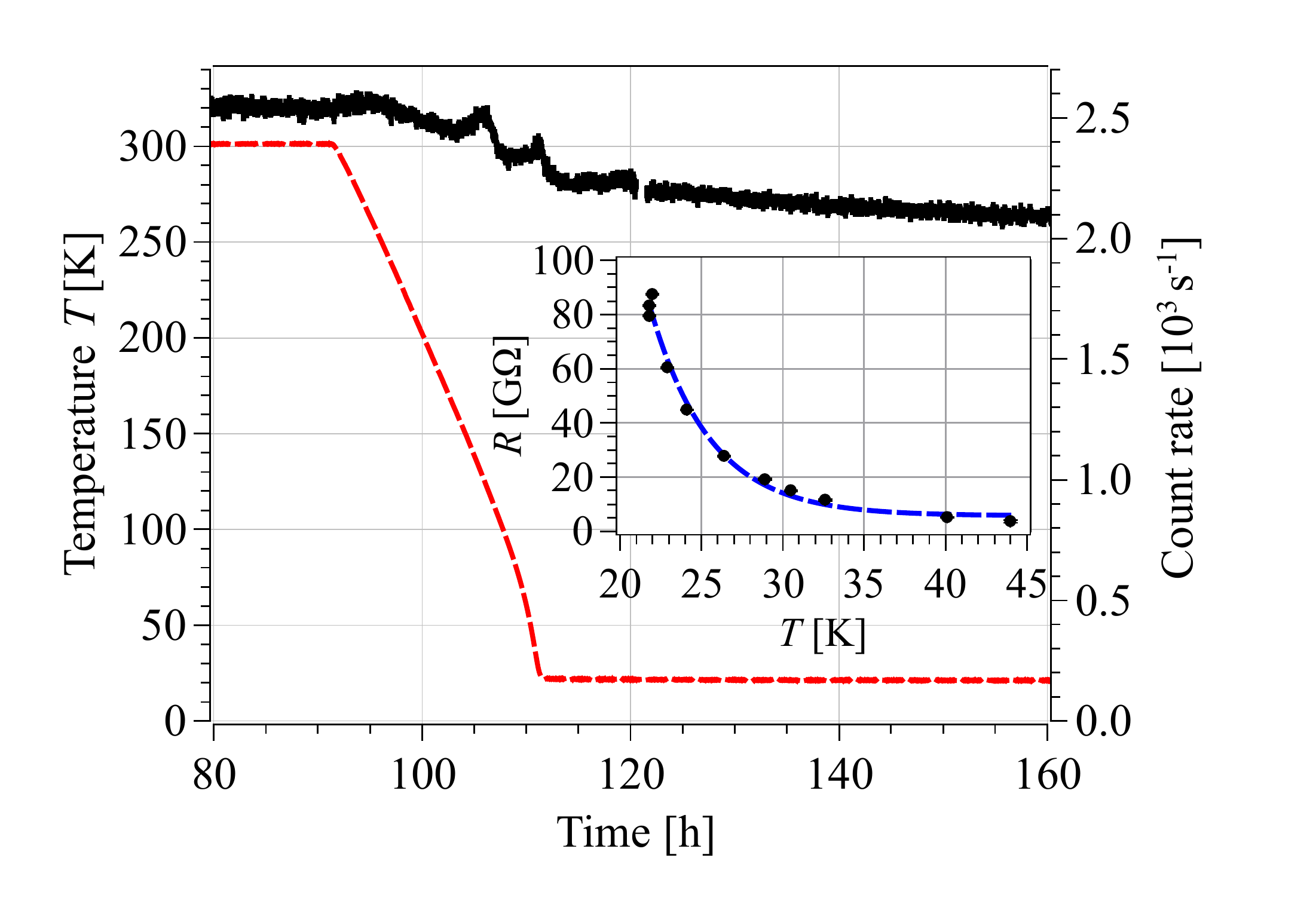}
\caption{\label{cooldown}(Colour online) Rate of discriminated alpha particle hits (black dots) during cool down of the sensor in the cryogenic test chamber. The horizontal axis shows the time since the beginning of the experiment, which lasted for approximately two weeks. The red dashed curve represents the measured temperature of the detector housing. The inset shows the resistance $R$ of the MCP stack in the low-temperature region. The blue dashed curve is an exponential fit to the data (black dots).}
\end{figure}  
\begin{figure}[tb]
\includegraphics[width=8cm]{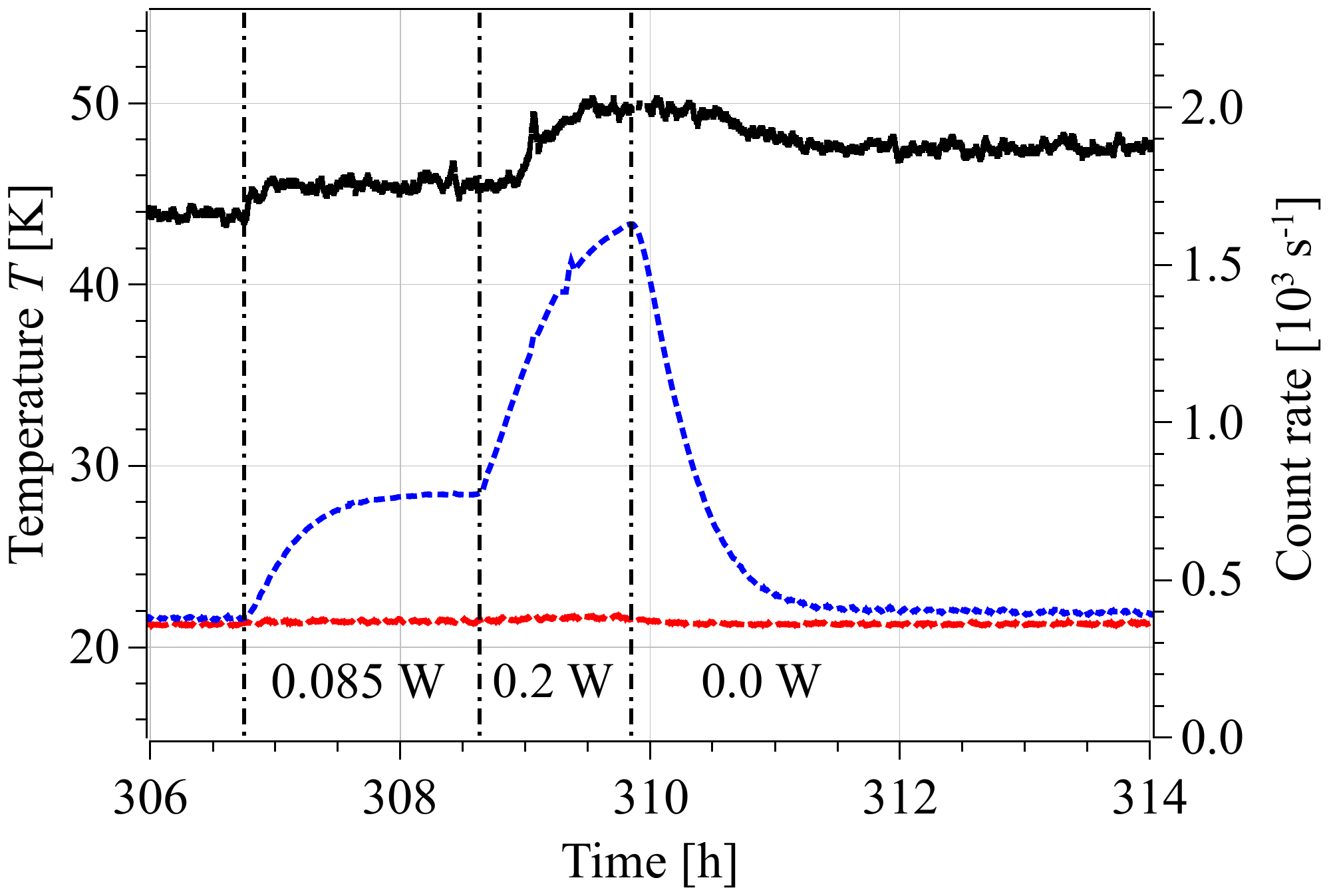}
\caption{\label{mcpheating}(Colour online) Effect of localised heating of the MCPs. The long-dashed red curve represents the temperature of the sensor housing. The short-dashed blue line is the temperature of the MCP set, obtained by calibration of the bias current. The black dots represent the discriminated alpha particle pulse rate. During the indicated time intervals, the MCP heating operated at powers of 85~mW and 200~mW, respectively.}
\end{figure}  
The evolution of the discriminated pulse rate during cool-down of the sensor from 300~K to 20~K is shown in Fig.~\ref{cooldown}. The temperatures in various places of the test cryostat were measured using Pt1000 thermo-sensors, one of which was attached to the outside of the grounded detector housing. During the cooling process, the measured count rate decreased by 8\%, from 2.5 to 2.3$\times 10^3$~s$^{-1}$. During the same time interval the resistance of the MCP stack, derived from the measured bias current at an operating voltage of 1750~V, rose from 56~M$\Omega$ at 300~K to 85~G$\Omega$ at 22~K. Hence, the initial decrease in the discriminated count rate can be interpreted as a reduction of the average pulse height due to the rise in MCP resistance. It was followed by a second, very slow, decrease of the discriminator rate at constant temperature that could be observed for several days of measurement (cf.\ Fig.~\ref{cooldown}). This cannot be explained by further increase of the MCP resistance, as the measured bias current was then constant. Instead, we believe that this behaviour is, at least partly, due to a variation in gain related to cryo-adsorption of residual gas onto the MCP. This explanation is supported by the response of the set-up to localised heating of the MCP as discussed below.\par     
Using a 5-W heater attached to the cold stage of the test cryostat, the temperature of the entire set-up could be varied between 22 and 45~K. In this way, the MCP bias current could be precisely calibrated against temperature in that range. In good approximation, we found the resistance of the channel plates to rise exponentially with decreasing temperature, as shown in the inset of Fig.~\ref{cooldown}, and as reported previously by other authors \cite{rosen, kuehnel,roth,schecker}.\par 
The behaviour of the MCPs below 22~K could not be probed due to the limited power of the cryo cooler and accuracy of the Pt1000 sensors. 
At CSR, significantly lower temperatures of $\sim 10$~K are expected, which will lead to even higher resistance of the channel plates than observed in this experiment. In order to guarantee good performance in CSR operation, a means to locally raise the MCP temperature relative to the cryogenic environment is required. This possibility is provided by the heating element included in the MCP-anode stack as described in Sect.~\ref{design}.\par
The response of the sensor to the MCP heating is shown in Fig.~\ref{mcpheating}. During the indicated time intervals, electric current was fed through the Constantan heating wire, first at a power of 85~mW, then of 200~mW. Using the MCP resistance as a proxy for temperature (cf.~Fig.~\ref{cooldown}) a prompt warming of the channel plates was observed after switching on the heating. The temperature would typically stabilise after $\sim 2$~h. For a heating power of 200~mW, the MCP temperature rose by more than 20~K, while the temperature of the detector housing was practically unaffected. This suggests that the heating is an efficient way of warming the MCP locally to a known-good operating temperature without disturbing the surrounding cryostat.\par 
No deterioration of the anode signal quality by the heating current was observed. The effect of the warming of the MCP on the pulse heights is shown in Fig.~\ref{mcpheating}: A significant rise in the discriminator rate was visible as soon as the heating element was switched on. Moreover, part of this rate enhancement was irreversible: Even after the heating was switched off and the MCP cooled down again, the discriminator rate remained at a $\sim 15$\% higher level than before the heating cycle. This supports our interpretation that the decrease in MCP pulse heights is partly due to residual gas adsorption onto the channel plates as noted above: As soon as the MCP is locally warmed, the surface contamination evaporates and is adsorbed onto the cold environment instead. At the CSR, which provides base vacuum conditions that are orders of magnitude better compared to the test cryostat, these effects are expected to be much less pronounced.
\subsubsection{Mechanics}
Using the same cryostat set-up, the mechanical components of the detector were tested for reliable operation at low temperature. The translation stage successfully performed 100 motion cycles over the full 300-mm stroke without apparent deterioration of the PEEK/stainless-steel worm drive or the flexible electric connections. Furthermore, independent tests of the reliability of the isolation vacuum drive and of the cryogenic rotary feed-through were performed at 38~K and 22~K, respectively. Although the manufacturer (\emph{Agilent Technologies}) specifies a minimum operating temperature of 77~K, the cold feed-through performed a total of 30\,000 revolutions at 22~K without resulting vacuum leak.
\subsubsection{Choice of electron multiplier}
As stated in Sect.\ \ref{design}, an EDR micro-channel plate was chosen as secondary electron multiplier stage for the detector, as its cryogenic behaviour could be expected to be favourable from previously published data. However the possibility to use a single-channel electron multiplier was also briefly investigated. CEMs provide higher gain and stronger pulse height saturation compared even to a chevron MCP stack. Hence they could be an interesting upgrade for future revisions of the detector. Two channel electron multipliers were tested: A standard, high-resistance ($\sim 400$~M$\Omega$ at 300~K) CEM by \emph{Philips} (type X 719 BL) and a lower-resistive ($\sim 70$~M$\Omega$) enhanced dynamic range CEM by \emph{Photonis} (type 7010M C EDR).\par
The CEMs were incorporated into a spare particle sensor as described by Rinn et al.\cite{rinn}, which operates on the same principle of secondary-electron ejection from a conversion electrode as the detector described in this work. The sensor was mounted onto the cold stage of the test cryostat and cooled to a temperature of $\sim 25$~K while being irradiated by the $^{241}$Am source as described above. It turned out that the EDR CEM did produce anode pulses at 25~K, although they were of much lower intensity than in room temperature operation. The loss in gain related to cooling was significantly stronger compared to the EDR MCP stack, but could be compensated by raising the operating voltage of the CEM from initially $1.8$~kV to $2.9$~kV. The standard resistance CEM, however did not produce measurable anode pulses at 25~K, although it operated normally before the cool-down and after rewarming to room temperature.\par
\begin{figure}[tb]
\includegraphics[width=8cm]{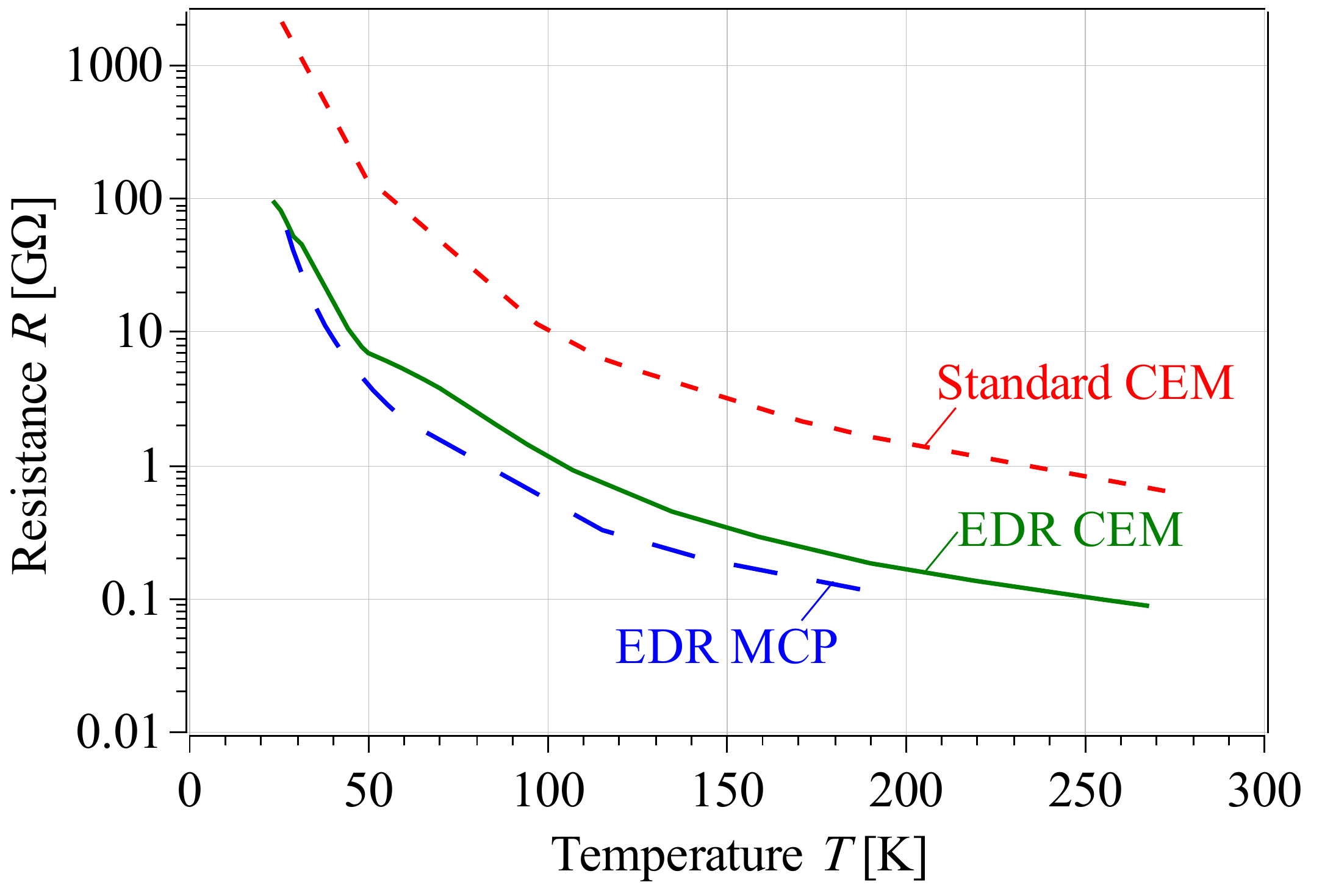}
\caption{\label{cemres}(Colour online) Electric resistance of the three tested electron multipliers, measured as a function of temperature. The EDR MCP (long-dashed blue curve) and EDR CEM (solid green) display approximately the same behaviour and both could be operated at 25~K. The standard CEM (short-dashed red) has got a resistance that is higher by a factor of $\sim 20$ across the entire temperature range and did not produce measurable electron pulses at 25~K (see text).}
\end{figure} 
Figure~\ref{cemres} shows the electric resistance of all three electron multipliers, measured using a high-voltage ohmmeter as a function of detector temperature in the range from 25 to 270~K. Both EDR MCP and EDR CEM display approximately the same behaviour of resistance versus temperature. The standard CEM, however, has a resistance that is larger by more than an order of magnitude compared to the EDR detectors, amounting to $\sim 3$~T$\Omega$ at 25~K. We suspect that this extremely high resistance limits the bias current of the CEM to the point where the channel wall can effectively not recharge such that sustained electron multiplication is not possible any more.\par
We note that operation of an EDR CEM in a cryogenic detector for the CSR storage ring seems possible, especially if the electron multiplier was equipped with a local heating element as described above. As all secondary electrons emitted from the conversion cathode enter the same amplification channel, a CEM has the potential of producing saturated, narrow pulse height distributions that can be easier to discriminate than the output signals of MCPs. However, more detailed tests of the CEM behaviour are necessary and the proven MCP stack was used as electron multiplier stage in this work.        
\section{Commissioning at the Storage Ring}
\label{commissioning}
The full detector set-up as shown in Fig.~\ref{CSRdetail} including all mechanics and electrical wiring was tested in operation during room-temperature commissioning \cite{grieseripac} of CSR in April 2014. For detector tests, the storage ring operated using a beam of 50-keV N$^+_2$, produced from the injection accelerator. Most of the CSR pumping facilities had still to be installed, and cryogenic operation was not yet foreseen. Due to the limited pumping speed, the residual gas pressure in the ring was not better than $\sim 1 \times 10^{-7}$~mbar, even after an intermediate bakeout at 120$^\circ$C had been performed. Consequently, the storage life time $\tau$ of the N$^+_2$ ion beam was $\sim$1~ms only.\par
\begin{figure}[tb]
\includegraphics[width=8cm]{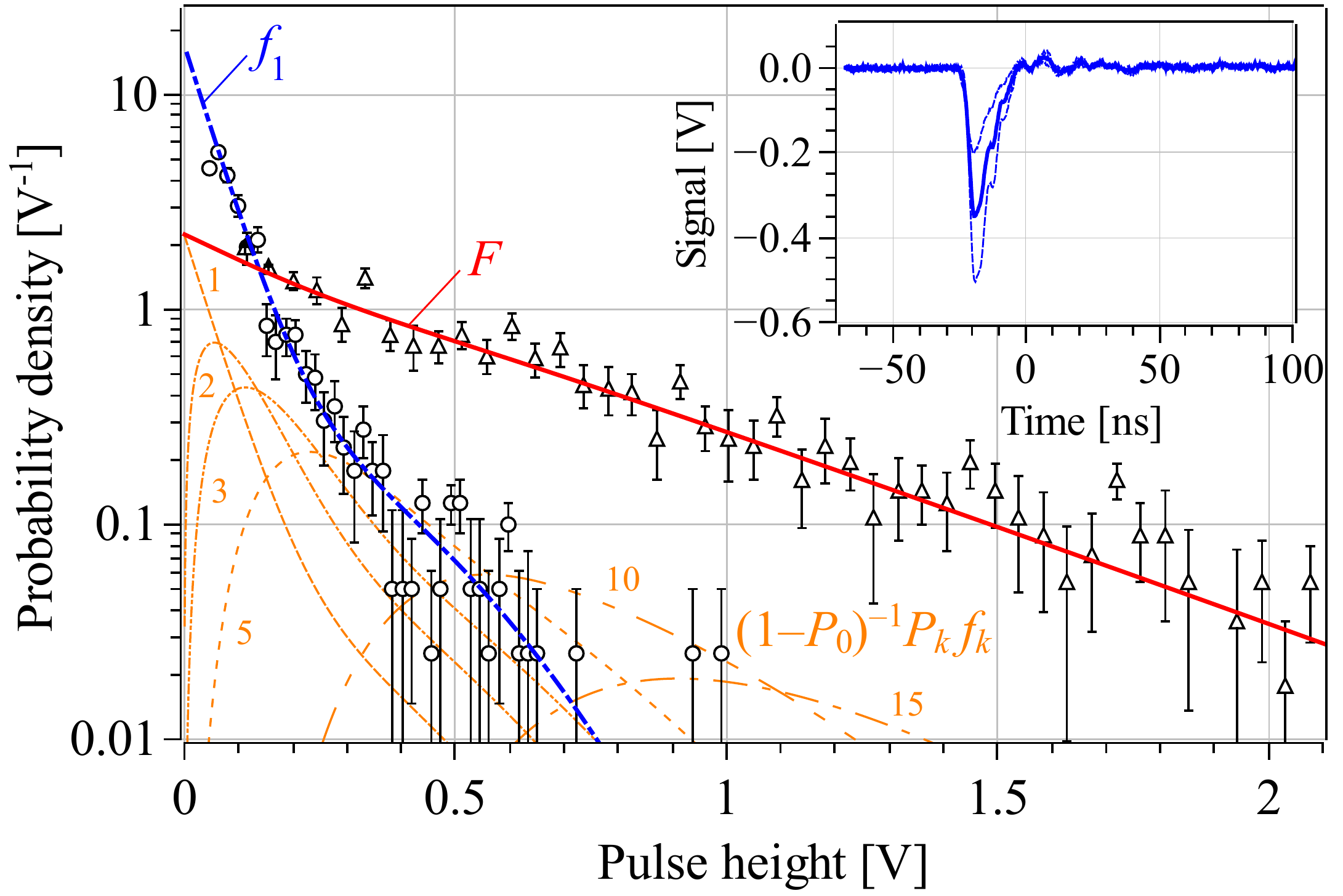}
\caption{\label{pulse}(Colour online) Measured pulse height spectra for UV irradiation (circles) and for 25-keV N$^+$ ions (triangles). The dashed blue curve ($f_1$) is a fit to the UV-induced single secondary electron spectrum. The solid red line ($F$) is modelled from the latter assuming Pólya-distributed secondary electron numbers $k$. Some individual contributions $P_kf_k$ are labelled 1 to 15 (see text). The inset displays an averaged anode pulse shape obtained for 25-keV N$^+$. The anode pulses were driven into 50~$\Omega$ and amplified by a factor 200 (cf.\ Sect.~\ref{design}).}
\end{figure}

By collisions with residual gas particles (X), the N$^+_2$ molecular ion most frequently dissociated into a neutral/charged fragment pair
\begin{equation}
 \mathrm{N}^+_2 + \mathrm{X} \rightarrow \mathrm{X} + \mathrm{N} + \mathrm{N}^+\ .
\end{equation}
The products N and N$^+$ had a kinetic energy of 25~keV each and were separated from the stored N$^+_2$ beam by the 6$^\circ$ deflector element preceding the detector in the beam line as shown in Fig.~\ref{CSRdetail}. The corresponding parameters $\eta$ from Eq.~(\ref{etaequation}) are $+1.0$ and $-1.0$ for the charged and neutral fragment, respectively. Hence, both daughters could be detected selectively by moving the sensor to the corresponding position in the CSR beam line. Using the electronics scheme described above, the anode pulses triggered by the impinging atomic particles were recorded. Despite the long and complicated electric line which connects the sensor anode to the decoupling and amplification circuit (cf.\ Sect.\ \ref{design}), necessarily involving several changes in impedance, the signal quality proved to be quite good. As shown in Fig.~\ref{pulse}, the anode pulses were followed by only little ringing, allowing for easy and efficient signal discrimination.\par
As the stored N$_2^+$ beam was uncooled, the N and N$^+$ products could be expected to irradiate the full sensitive area of the sensor. In contrast to Sect.\ \ref{testpulse}, the number $k$ of secondary electrons multiplied in the MCPs is then gouverned by Pólya statistics according to Eq.~(\ref{pulseheightpolya}). The pulse height distribution is defined by the electric response of the MCP stack to those $k$ electron avalanches.\par 
An ultraviolet (UV) AlGaN light emitting diode (LED, \emph{SETi} UVTOP\,240) is installed in the CSR for detector testing purposes. It allows to irradiate particle detectors with an uncollimated beam of $\sim 245$-nm photons. Their energy of 5.1\,(1)~eV allows the photons to eject a single electron \emph{at maximum} upon absorption at the conversion cathode. The UV LED hence provides a means to relate the measured pulse height distribution for heavy particle impact to the average number $\tilde{\gamma}$ of secondary electrons reaching the MCP.\par
The pulse height distribution $f_1$ measured for UV photon irradiation is shown in Fig.~\ref{pulse}. The MCP stack operated at a voltage of 950~V per plate. If $f_1$ is the pulse height spectrum corresponding to precisely one initial electron multiplied at the MCP set, then the response for precisely two electron cascades is given by the convolution $f_2 = f_1 * f_1$, and for precisely $k$ cascades by $f_k = f_1 * f_{k-1}$. The pulse height spectrum $F$ for N$^+$ ions can thus be modelled from the measured single-electron detector response $f_1$ as 
\begin{equation}
 \label{Fdistr}
 F= (1-P_0)^{-1} \times \sum_{k=1}^\infty P_k f_k\ .
\end{equation}
The weights $P_k$ are taken from  Eq.~(\ref{pulseheightpolya}) as a function of the parameters $\tilde{\gamma}$ and $b$. The correction factor $(1-P_0)^{-1}$ ensures that the resulting distribution $F$ is normalised to 1, although there is a non-vanishing probability ($P_0$) for an ion to produce no signal.\par
Figure \ref{pulse} shows the measured pulse height spectrum for N$^+$ products impinging onto the detector, along with a model spectrum $F$ derived from the measured UV-photon response $f_1$. Empirically, the UV response was found to be well reproduced by a sum of an exponential decay and a Gaussian centred at zero. Subsequently, the convolutions $f_n$ were computed numerically. The best fit for $F$ from Eq.\ (\ref{Fdistr}) to the N$^+$ data is obtained for $\tilde{\gamma} = 10(1)$ and $b=0.8(2)$ (cf.\ Eq.~(\ref{pulseheightpolya})). The fit model includes no other free parameters. As a large area of the converter cathode was irradiated in this experiment, a large value of $b$ was to be expected. Since the Pólya parameter $b$ is the relative variance\cite{prescott} of the distribution of $\gamma \epsilon_\text{c}$, this corresponds roughly to values $\gamma \epsilon_\text{c} \approx 10 \pm 9$ across the entire converter cathode area for 25-keV N$^+$ particles. Note however that the fit function $F$ depends quite strongly on the low-energy part ($\leq 50$~mV) of $f_1$ which, as visible in Fig.~\ref{pulse}, has not been measured but extrapolated. Hence, this result is likely bound to a large systematic uncertainty. Still, it seems that a basic understanding of the pulse height spectra can be obtained from Eq.~(\ref{pulseheightpolya}). The measured pulse height distribution for 25-keV N$^+$ also shows that the loss in detection efficiency by the discrimination threshold -- which can be as low as a few 10~mV, thanks to the good signal quality -- is small (a few 10$^{-2}$).\par 
\begin{figure}[tb]
\includegraphics[width=8cm]{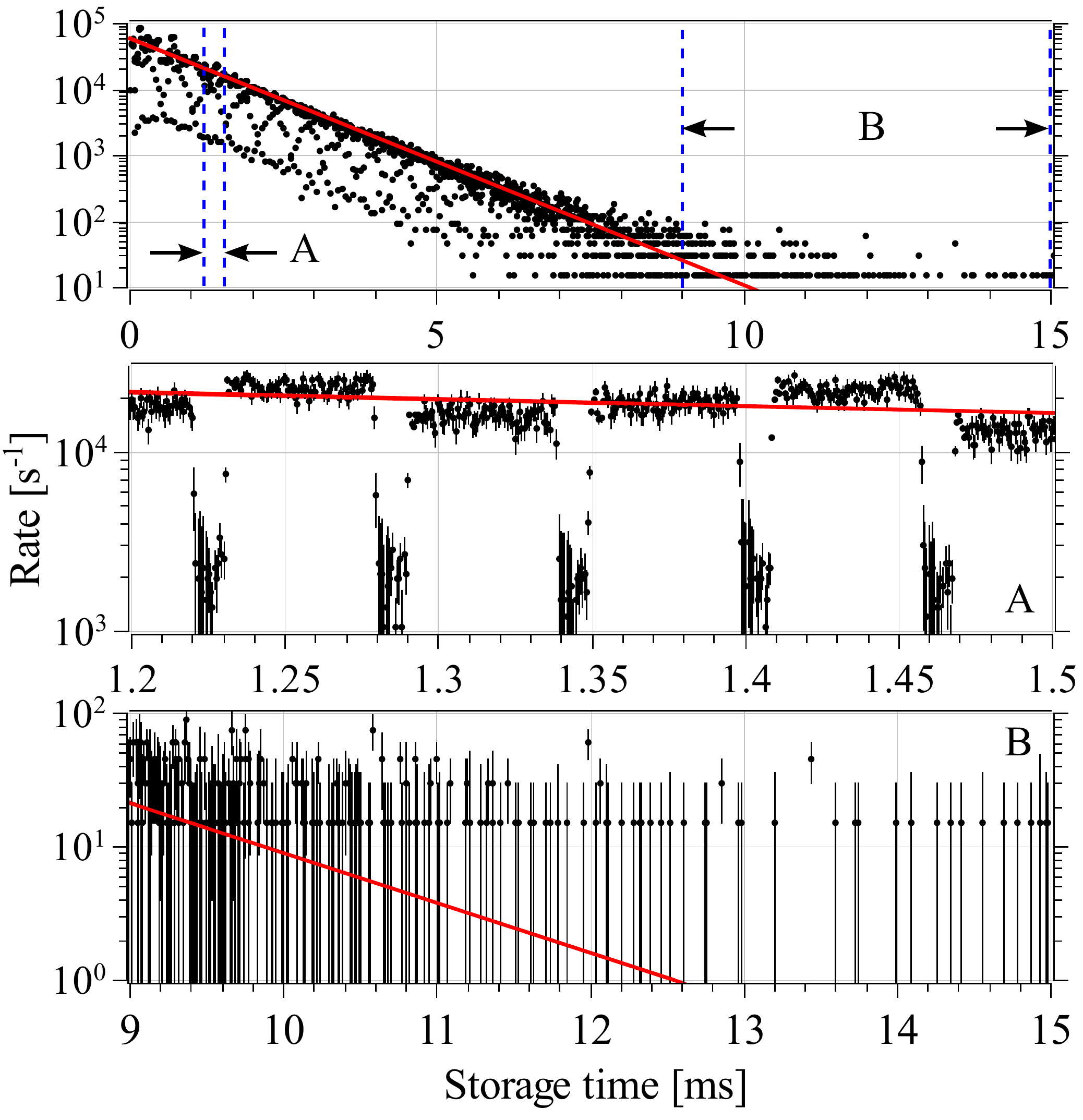}
\caption{\label{rates}(Colour online) Measured rate of N$^+$ impacts (black dots) onto the detector as a function of storage time of the N$^+_2$ beam in CSR. The red line indicates the exponential decay of the ion beam with lifetime $\tau = 1.15(5)$~ms. The second and third frames, labelled A and B, are close-ups of the signal as indicated by the dashed blue lines.}
\end{figure}
Figure~\ref{rates} shows the measured detector count rate as a function of storage time of the N$^+_2$ beam in CSR. The data is averaged over several thousand individual ion injections into the storage ring. The sharp modulation of the signal, clearly visible in frame A of Fig.~\ref{rates}, stems from the fact that the injected ion bunch had a length of only 29~m, as opposed to the full closed orbit circumference of CSR which is of 35.4~m. Hence, the train of particles passed by the detector only for $\sim 80$\% of the time of one revolution period, which was of 59.4~$\mu$s. For both, neutral and charged products, count rates up to several 10$^4$~s$^{-1}$ were recorded directly after ion injection, followed by an exponential decrease with decay constant $\tau=1.15(5)$~ms, which matched the anticipated lifetime of the ions due to residual gas collisions. After $\sim 15$~ms, corresponding to $\sim 250$ revolutions of the ion bunch in the storage ring, the measured product rate vanished, showing that the EDR MCP stack is effectively free of any dark count rate, even in room-temperature operation.          
\section{Conclusion}
\label{summary}
We have designed a highly efficient, movable single-particle counter that is suitable for operation in a cryogenic ultra-high vacuum environment. While the device has been developed specifically for operation at the CSR storage ring, we note that the latter shares many of its characteristics with similar electrostatic and cryogenic storage devices in operation\cite{desiree} or in construction\cite{riken}. Hence, we believe that our development work is of interest also to others.\par 
The detector has been successfully commissioned at CSR in room-temperature operation. The set-up is bakeable to high (250$^\circ$C) temperature, and off-line performance tests have shown that it is fully suited for cryogenic operation in the $\sim 10$-K environment of upcoming CSR experiments. The particle sensor geometry has been optimised using simulations of the secondary-electron optics, and direct measurements have confirmed an overall detection efficiency close to 100\% for particle energies of a few keV/u.\par
While the set-up has been designed primarily with future electron cooler experiments in mind, we expect to develop variants of the detector also for other experimental sections of the CSR storage ring. In that process, also the possibility of using a CEM as alternative electron multiplication stage will be further investigated.   
\section*{Acknowledgements}
\label{acknowledgements}
We gratefully acknowledge the technical support this work has obtained from the MPIK mechanical workshop and accelerator staff as well as the financial support it has received from the Max Planck Society (MPG). O.~N. was supported, in part, by the United States National Science Foundation (NSF) and National Aeronautics and Space Administration (NASA). The work of A.~B. as well as that of K.~S. was funded by the German Research Foundation (DFG) within the Priority Programme 1573, under contract numbers Wo 1481/2-1 and Schi 378/9-1, respectively.\par

\end{document}